\documentclass[aps,prd,reprint,groupeaddress,showpacs,preprintnumbers,nofootinbib,nobibnotes,amsmath,amssymb,floatfix,superscriptaddress]{revtex4-2}
\usepackage[colorlinks=true,linkcolor=red,citecolor=blue,urlcolor=blue]{hyperref}
\usepackage[version=4]{mhchem}
\usepackage{hyperref}
\hypersetup{
  colorlinks   = true,    % Colours links instead of ugly boxes
  urlcolor     = blue,    % Colour for external hyperlinks
  linkcolor    = navy,    % Colour of internal links
}

\usepackage{booktabs}

\usepackage{multirow}
\usepackage{amsmath} 
\usepackage{physics}
\usepackage{graphicx}
\usepackage{color}
\usepackage{float}
\usepackage{xcolor}
\usepackage{amssymb}
\usepackage{enumitem}
\usepackage{caption}
\usepackage{natbib}
\usepackage[T1]{fontenc}
\definecolor{red}{rgb}{0.8,0,0}
\definecolor{violet}{rgb}{0.4,0,0.4}
\definecolor{green}{rgb}{0,0.5,0.0}
\definecolor{navy}{rgb}{0.0,0.0,0.6}
\definecolor{orange}{rgb}{0.8,0.2,0.0}
\usepackage[normalem]{ulem}  % \sout{old text} for strikeout

\newcommand{\apjl}{Astro. Phys. J. Lett.}

\newcommand{\aap}{Astronomy $\&$ Astrophysics}
\newcommand{\nphysa}{Nuc. Phys. A}

\newcommand{\sovast}{Sov. Astron.}

\begin{document} 

\title{(Anti)kaon condensation in strongly magnetized dense matter}

\author{Debraj Kundu}
\email{kundu.1@iitj.ac.in}
\affiliation{Indian Institute of Technology Jodhpur, Jodhpur 342037, India}
\author{Vivek Baruah Thapa}
\email{thapa.1@iitj.ac.in}
\affiliation{Indian Institute of Technology Jodhpur, Jodhpur 342037, India}
\affiliation{National Institute of Physics and Nuclear Engineering (IFIN-HH), RO-077125, Bucharest, Romania}
\author{Monika Sinha}
\email{ms@iitj.ac.in}
\affiliation{Indian Institute of Technology Jodhpur, Jodhpur 342037, India}

\date{\today}
\begin{abstract}
{Recent observations of several massive pulsars, with masses near and above $2~M_\odot$, point towards the existence of matter at very high densities, compared to normal matter that we are familiar with in our terrestrial world. This leads to the possibility of appearance of exotic degrees of freedom other than nucleons inside the core of the neutrons stars (NS). Another significant property of NSs is the presence of high surface magnetic field, with highest range of the order of $\sim~10^{16}$ G. We study the properties of highly dense matter with the possibility of appearance of heavier strange and non-strange baryons, and kaons in presence of strong magnetic field. We find that the presence of a strong magnetic field stiffens the matter at high density, delaying the kaon appearance and, hence, increasing the maximum attainable mass of NS family. 
}
\end{abstract}

\maketitle

\section{Introduction}\label{sec 1}

The state of matter inside neutron stars (NSs) is an unsolved mystery of modern science. Born from the remnants of a supernova explosion, a neutron star exhibits a range of densities inside its structure, the density \textcolor{black}{at} the core possibly being several times that of \textcolor{black}{nuclear saturation density \cite{glendenning2012compact,WEBER200794, SEDRAKIAN2007168, LATTIMER2016127, 2021ARNPS..71..433L, 2000ARNPS..50..481H}}. Many recent astrophysical observations indicate that the possible lower limit of NS maximum mass is above $2~M_\odot$, viz. PSR J1614-2230 ($M= 1.97 \pm 0.04 M_{\odot}$) \cite{_zel_2010,2010Natur.467.1081D}, MSP J0740+6620 ($M= 2.14^{+0.20}_{-0.18} M_{\odot}$ with 95\% credibility) \cite{2020NatAs...4...72C}, PSR J0348+0432 ($M= 2.01 \pm 0.04 M_{\odot}$) \cite{2013Sci...340..448A} and PSR J0952-0607 ($M= 2.35 \pm 0.17 M_{\odot}$) \cite{2022ApJ...934L..17R}. These findings strengthen the idea of \textcolor{black}{the} existence of highly dense matter in the core of NSs. Thus, investigation of the matter inside NSs provides us with \textcolor{black}{a} unique opportunity to study matter under extreme conditions that cannot be \textcolor{black}{attained in any of the terrestrial} laboratories.

The gravitational pull inside the NS is balanced mostly by the Fermi degeneracy pressure of neutrons, along with some amounts of protons and leptons (electrons and muons). In addition, the extreme matter density inside NSs can lead to energetically favorable conditions for exotic particles to appear. Hyperons are one such species of particles that might appear inside the NS if the baryon chemical potential becomes high enough. The possibility of their occurrence was first suggested in \cite{1960SvA.....4..187A}. Another class of particle species that might make its appearance is $\Delta$-resonances. Its appearance pushes the threshold for \textcolor{black}{the} onset of hyperons to higher densities \cite{LI2018234,Li_2019,PhysRevD.89.043014}.

Similarly, another possible addition to the degrees of freedom can come from the appearance of meson condensates \cite{2015arXiv150705839P} if the lepton chemical potential becomes high enough. However, for the lowest massive meson $\pi$ \textcolor{black}{(pion)}, the repulsive $s$-wave pion-nucleon scattering potential increases the effective ground state mass of $\pi$-meson \cite{1985ApJ...293..470G, PhysRevC.80.038202}.
\textcolor{black}{However, a few works \cite{Ericson:1988gk, 2019Symm...11..778K} have argued the possibility of pion condensation due to the fact that $p-$wave scattering potential is attractive in nature.}
%Hence, the possibility of $\pi$-meson's appearance is ruled out.
On the other hand,
%for the next massive $K$-mesons,
(anti)kaon ($\bar{K}\equiv K^{-} , \bar{K}^{0}$) mesons may appear in the form of s-wave Bose condensates due to the attractive nature of (anti)kaon optical potential. $K^{+}$ and $K^0$ kaons have repulsive optical potentials and their presence in nuclear matter increases their effective masses. Thus, the occurrence of $K^{+}$ and $K^0$ in NS matter is discouraged. The threshold density for the onset of $\bar{K}$ is highly sensitive to its optical potential and whether or not hyperons are present \cite{PhysRevC.90.015801}. The presence of $\bar{K}$ in NS matter has been extensively studied \textcolor{black}{in past literature }\cite{PhysRevC.60.025803,PhysRevC.63.035802,PRAKASH19971,PhysRevC.53.1416,PhysRevC.64.055805,Malik2021NewEO,PhysRevD.102.123007}.

As already mentioned, the verification of the theoretical models of highly dense matter can only be done with the observations from NSs. The astrophysical observable properties of NSs %constructed by the dense matter models
should be studied to constrain the \textcolor{black}{dense matter models}. For example, one should note that the appearance of hyperons tends to soften the equation of state (EoS) and, consequently, results in lowering of the maximum mass of NSs. Studies \cite{LI2018234,Li_2019} have indicated that the inclusion of $\Delta$-resonances does not affect the implied maximum mass significantly, but it reduces the radius and thereby increases the compactness of the stars. The appearance of $\bar{K}$, similar to hyperons, softens the EoS and, thus, lowers the maximum mass of NSs.

The theoretical model of dense matter can be obtained from terrestrial laboratory data by extrapolating the nuclear matter properties at nuclear saturation density and it can be further constrained from the recent mass-radius measurements of NSs, viz. the NICER mission observations give the mass-radius measurements of PSR J0030+0451 as $1.44^{+0.15}_{-0.14} \; M_{\odot}$, $13.02^{+1.24}_{1.06}$ km \cite{Miller_2019} and $1.34^{+15}_{-16} \; M_{\odot}$, $12.71^{+1.14}_{-1.19}$ km \cite{Riley_2019}, respectively. Another important constraint on highly dense matter inside NSs comes from the gravitational wave detection observations which provide us with the estimate of maximum limit of tidal deformability of the star made of highly dense matter.

Another salient feature of NSs is their strong surface magnetic field in the range $10^8-10^{16}$ G. A particular class of NSs which have ultra strong surface magnetic field of $10^{14} - 10^{16}$ G \cite{Harding_2006,Turolla_2015}  are called magnetars. The matter inside NSs also experiences Pauli paramagnetism and Landau diamagnetism. Pauli paramagnetism is applicable for both charged and uncharged particles while the Landau diamagnetism affects only charged particles, being particularly strong for light particles like leptons. In our present work, we first note down the constraint on model parametrizations from the astrophysical observations of mass-radius measurements of many pulsars as well as tidal deformability from GW observations. Then, with the constrained model, we study the properties of dense matter and NSs with strong magnetic field. 

Previous studies have been conducted on NS matter containing hyperons and $\Delta$-resonances without (anti)kaon condensates \cite{LI2018234,Li_2019} and with (anti)kaon condensates \cite{PhysRevD.103.063004}. The study without (anti)kaons was also extended to accommodate strong magnetic fields in \cite{particles3040043}. Work has also been done on NS matter containing (anti)kaon condensates, but no hyperons or $\Delta$-resonances, under the effect of strong magnetic fields \cite{PhysRevC.77.045804,2022IJMPE..3150050K}. In this paper, we present the novel study of matter inside NS having a strong magnetic field (magnetar) containing hyperons, (anti)kaon condensates and $\Delta$-resonances (as exotic degrees of freedom) in $\beta$-equilibrium. We have used the relativistic mean field (RMF) model to describe the interactions between the particles. As the soft matter with hyperons attains the lower limit of maximum mass with density dependent baryon-meson interactions, we use density dependent RMF (DD-RMF) model to study the effect of strong magnetic field on NS composed of matter with (anti)kaon condensates along with $\Delta$-resonances and hyperons.

In the next section (sec. \ref{sec 2}) we discuss the matter model under the effect of magnetic field. Then, in sec. \ref{Sec 3}, we discuss the results with model parameters compatible with the the astrophysical observations.
Section \ref{sec:summary} presents a brief summary of our work.

\section{Formalism}\label{sec 2}

\subsection{DD-RMF Model}

Here, we lay down the formulation for the DD-RMF model. We consider the NS matter to be composed of nucleons ($n, p$), leptons ($e^-, \mu^-$), hyperons ($\Lambda, \Xi, \Sigma$), (anti)kaons ($\bar{K}\equiv K^{-} , \bar{K}^0$) and $\Delta$-resonances ($\Delta^{-}, \Delta^{0}, \Delta^{+}, \Delta^{++}$). In this model, the strong interactions between the nucleons, hyperons, (anti)kaons and $\Delta$-resonances are mediated by the following meson fields: isoscalar-scalar $\sigma$, isoscalar-vector $\omega^{\mu}$ and isovector-vector $\boldsymbol{\rho}^{\mu}$. We have also considered the strange isoscalar-vector meson field $\phi^\mu$ as a mediator of hyperon-hyperon and (anti)kaon-hyperon interactions. Throughout our work, we have used the natural units, $\hbar = c = G = 1$.

The total Lagrangian density is given by: \cite{LI2018234,PhysRevC.60.025803,PhysRevC.63.035802,PAL2000553,Broderick_2000,PhysRevC.91.045803,PhysRevC.89.015805, PhysRevC.77.045804}

\begin{equation}\label{1}
    \mathcal{L}= \mathcal{L}_m + \mathcal{L}_{em}
\end{equation}
where $\mathcal{L}_m$ and $\mathcal{L}_{em}$ are the matter and the electro-magnetic field contributions, respectively. 

For the matter part of the Lagrangian density, We have

\begin{align}\label{2}
\begin{split}
    \mathcal{L}_m &= \sum_b\bar{\psi}_b(i\gamma_\mu D^\mu_{(b)} - m^*_b)\psi_b + \sum_d\bar{\psi}_{d\nu}(i\gamma_\mu D^\mu_{(d)} - m^*_d)\psi^\nu_d \\
    &+ \sum_l\bar{\psi}_l(i\gamma_\mu D^\mu_{(l)} - m_l)\psi_l+D_\mu^{(\bar{K})*}\bar{K}D^\mu_{(\bar{K})}K - m^{*2}_K\bar{K}K\\
    &+\frac{1}{2}(\partial_\mu\sigma\partial^\mu\sigma - m_\sigma^2\sigma^2) - \frac{1}{4}\omega_{\mu\nu}\omega^{\mu\nu} +  \frac{1}{2}m_\omega^2\omega_\mu\omega^\mu \\
    &- \frac{1}{4}\boldsymbol{\rho}_{\mu\nu}\cdot\boldsymbol{\rho}^{\mu\nu} + \frac{1}{2}m_\rho^2\boldsymbol{\rho}_\mu\cdot\boldsymbol{\rho}^\mu - \frac{1}{4}\phi_{\mu\nu}\phi^{\mu\nu} +\frac{1}{2}m_\phi^2\phi_\mu\phi^\mu
\end{split}
\end{align}
where $\psi_b$, $\psi_d^\nu$  and $\psi_l$ represent the fields of octet baryons, $\Delta$-resonances and leptons, respectively.
\textcolor{black}{And $\bar{K}$ represents the (anti)kaon condensate fields.}
\textcolor{black}{$\Delta-$resonances, being spin-3/2 particles, are governed by the Schwinger-Rarita field equations \cite{PhysRev.60.61}.}
$m_b$, $m_d$, $m_l$ and $m_K$ stand for the masses of octet baryons, $\Delta$-resonances, leptons and (anti)kaons, respectively. $\sigma$, $\omega_\mu$, $\boldsymbol{\rho}_\mu$ and $\phi_\mu$ are the meson fields with masses $m_\sigma$, $m_\omega$ , $m_\rho$ and $m_\phi$, respectively.
The covariant derivatives in Eqn.(\ref{2}) are given by:
\begin{align}\label{3}
\begin{split}
D_{\mu(j)}=& \partial_\mu + ig_{\omega j}\omega_\mu + ig_{\rho j}\boldsymbol{\tau}_j.\boldsymbol{\rho}_\mu + ig_{\phi j}\phi_\mu + ieQA_{\mu}\\
D_{\mu(l)}=& \partial_\mu + ieQA_{\mu}
\end{split}
\end{align}
with $j$ representing the octet baryons $(b)$, $\Delta$-resonances $(d)$ and (anti)kaons ($\bar{K}$), and $l$ representing leptons. $\boldsymbol{\tau}_j$ is the isospin operator for the $\boldsymbol{\rho}^\mu$ meson fields. $eQ$ is the charge of the particle with $e$ being unit positive charge. We choose the direction of magnetic field as the $z$-axis with the field four-vector potential as $A^\mu\equiv(0,-yB,0,0)$, with $B$ being the magnetic field magnitude. Under the effect of this magnetic field, the motion of the charged particles is Landau quantized in the plane perpendicular to the direction of field, the momentum in the perpendicular direction being $p_\perp = 2\nu e|Q|B$, where $\nu$ is the Landau level. Here, the baryon-meson coupling parameters are considered density dependent.

The gauge mesonic contributions in Eqn.(\ref{2}) contain the field strength tensors:
\begin{align}\label{4}
\begin{split}
    \omega_{\mu\nu}= \partial_\nu\omega_\mu - \partial_\mu\omega_\nu\\
    \boldsymbol{\rho}_{\mu\nu}= \partial_\nu\boldsymbol{\rho}_\mu - \partial_\mu\boldsymbol{\rho}_\nu\\
    \phi_{\mu\nu}= \partial_\nu\phi_\mu - \partial_\mu\phi_\nu
\end{split}    
\end{align}

The effective masses of the baryons and (anti)kaons used in Eqn.(\ref{2}) are given by:
\begin{align}\label{5}
\begin{split}
    m_b^*&= m_b - g_{\sigma b}\sigma\\
    m_d^*&= m_d - g_{\sigma d}\sigma\\
    m_K^*&= m_K - g_{\sigma K}\sigma
\end{split}    
\end{align}
$g_{\sigma j}$ in Eqn.(\ref{5}) and $g_{\omega j}$, $g_{\rho j}$, $g_{\phi j}$ in Eqn.(\ref{3}) are density dependent coupling parameters.

The electro-magnetic field part of the Lagrangian density in Eqn.(\ref{1}) is given by:
\begin{equation}
    \mathcal{L}_{em}= -\frac{1}{16\pi}F_{\mu\nu}F^{\mu\nu}
\end{equation}
where $F_{\mu\nu}$ is the electro-magnetic field tensor.\vspace{4pt}

In the relativistic mean field approximation, the meson fields acquire the following ground state expectation values:
\begin{equation}
    \begin{aligned}
        \sigma =&\sum_b\frac{1}{m_\sigma^2}g_{\sigma b}n_b^s + \sum_d\frac{1}{m_\sigma^2}g_{\sigma d}n_d^s + \sum_{\bar{K}}\frac{1}{m_\sigma^2}g_{\sigma K}n_{\bar{K}}^s\\
    \omega_0 = &\sum_b \frac{1}{m_\omega^2} g_{\omega b}n_b + \sum_d \frac{1}{m_{\omega}^2} g_{\omega d} n_d - 
    \sum_{\bar{K}} \frac{1}{m_{\omega}^2}g_{\omega K} n_{\bar{K}} \\
    \phi_0 =&\sum_b\frac{1}{m_\phi^2}g_{\phi b}n_b - \sum_{\bar{K}}\frac{1}{m_\phi^2}g_{\phi K}n_{\bar{K}}\\
    \rho_{03} =&\sum_b\frac{1}{m_\rho^2}g_{\rho b}\boldsymbol{\tau}_{b3}n_b + \sum_d\frac{1}{m_\rho^2}g_{\rho d}\boldsymbol{\tau}_{d3}n_d\\ &+ \sum_{\bar{K}}\frac{1}{m_\rho^2}g_{\rho K}\boldsymbol{\tau}_{\bar{K}3}n_{\bar{K}}
    \end{aligned}
\end{equation}
where the scalar density $n^s_j= \langle\bar{\psi}\psi\rangle $ and the vector (baryon) number density $n_j= \langle\bar{\psi}\gamma^0\psi\rangle$ .

The scalar density, baryon number density and the kinetic energy density of the uncharged baryons at the temperature $T=0$ limit are given by:

\begin{equation}
\begin{aligned}
       n^s_u =& \frac{2J_u+1}{2\pi^2}
       m^*_u\left[p_{F_u}E_{F_u}-m_u^{*^2}\ln{\left(\frac{p_{F_u}+E_{F_u}}{m_u^*}\right)}\right] \label{8} \\
    n_u =& (2J_u+1)\frac{p^3_{F_u}}{6\pi^2}\\
    \varepsilon_u =&  \frac{2J_u+1}{2\pi^2}
    \bigg[p_{F_u}E_{F_u}^3-\frac{m_u^{*^2}}{8}\bigg(p_{F_u}E_{F_u}\\ 
%    \left. \\ & \right. \\
    &+ m_u^{*^2}\ln{\bigg(\frac{p_{F_u} + E_{F_u}}{m_u^*}\bigg)}\bigg)\bigg] 
\end{aligned}
\end{equation}

where $J$, $p_F$ and $E_F$ represent the spin, Fermi momentum and Fermi energy, respectively. Here the uncharged baryons are denoted by subscript $u$.

The scalar density, baryon number density and the kinetic energy density of the charged baryons at the temperature $T=0$ limit are given by:
\begin{itemize}
\item For spin- 1/2 baryons:
\begin{align}
    n^s_c =& \; \frac{e|Q|B}{2\pi^2} m^*_c\sum^{\nu_{max}}_{\nu=0}(2-\delta_{\nu,0})\ln{\left(\frac{p_c(\nu) +E_{F_c}}{\sqrt{m_c^{*^2}+ 2\nu e|Q|B}}\right)}\\
    n_c =& \; \frac{e|Q|B}{2\pi^2} \sum^{\nu_{max}}_{\nu=0}(2-\delta_{\nu,0})p_c(\nu)\\
    \varepsilon_c =& \; \frac{e|Q|B}{4\pi^2} \sum^{\nu_{max}}_{\nu=0}(2-\delta_{\nu,0})\Bigg[p_c(\nu)E_{F_c} +  \Big(m_c^{*^2} + 2\nu e|Q|B\Big) \nonumber \\&   \ln{\left(\frac{p_c(\nu) +E_{F_c}}{\sqrt{m_c^{*^2}+ 2\nu e|Q|B}}\right)} \Bigg]
\end{align}
\item For spin- 3/2 baryons:
\begin{align}
    n^s_c =& \; \frac{e|Q|B}{2\pi^2} m^*_c\sum^{\nu_{max}}_{\nu=0}(4-\delta_{\nu,1} -2\delta_{\nu,0}) \nonumber \\
    &\ln{\left(\frac{p_c(\nu)+E_{F_c}}{\sqrt{m_c^{*^2}+ 2\nu e|Q|B}}\right)}\\
    n_c =& \; \frac{e|Q|B}{2\pi^2} \sum^{\nu_{max}}_{\nu=0}(4-\delta_{\nu,1} -2\delta_{\nu,0})p_c(\nu)\\
    \varepsilon_c =& \; \frac{e|Q|B}{4\pi^2} \sum^{\nu_{max}}_{\nu=0}(4-\delta_{\nu,1} -2\delta_{\nu,0})\bigg[p_c(\nu)E_{F_c} + \nonumber \\&\Big(m_c^{*^2} + 2\nu e|Q|B\Big) \ln{\left(\frac{p_c(\nu)+E_{F_c}}{\sqrt{m_c^{*^2}+ 2\nu e|Q|B}}\right)}\bigg]
\end{align}
\end{itemize}
where $p(\nu)= \sqrt{p_{F}- 2\nu eB}$. The charged baryons are denoted by subscript $c$ . The maximum value of $\nu$ is given by:
\begin{equation}\label{eqn. 15}
    \nu_{max}= \text{Int}\left(\frac{p_F^2}{2e|Q|B}\right).
\end{equation}
\textcolor{black}{In case of Dirac particles, the degeneracy of the lowest Landau level is unity and 2 for all other levels \cite{2013NuPhA.898...43S}. While for the Schwinger-Rarita particles, the same is 2 for the lowest, 3 in the second and 4 for the other remaining Landau levels \cite{2013JPhG...40e5007D}.}

The number density of (anti)kaon ($\bar{K}$) condensates is given by \cite{PhysRevC.77.045804}:
\begin{align}
    n_{K^-}=& \; 2\sqrt{m_K^{*^2} + |q_{K^-}|B} \: \bar{K}K \\
    n_{\bar{K}^0}=& \; 2m_K^*\bar{K}K
\end{align}
where $|q_{K^-}|$ is the charge of $K^-$.

In the case of leptons, the number density and kinetic energy density are given by:
\begin{align}
    n_l =& \; \frac{e|Q|B}{2\pi^2} \sum^{\nu_{max}}_{\nu=0}(2-\delta_{\nu,0})p_l(\nu)\\
    \varepsilon_l =& \; \frac{e|Q|B}{4\pi^2} \sum^{\nu_{max}}_{\nu=0}(2-\delta_{\nu,0})\bigg[p_l(\nu)E_{F_l} + \Big(m_l^2 + 2\nu e|Q|B\Big) \nonumber \\
    &\ln{\left(\frac{p_l(\nu)+E_{F_c}}{\sqrt{m_l^2+ 2\nu e|Q|B}}\right)}\bigg] \label{19}
\end{align}
 The leptons are denoted by subscript $l$. Throughout Eqns.\eqref{8} - \eqref{19}, we have
 \begin{equation}
     p_{F}= \sqrt{E_F^2 - m^{*^2}}
 \end{equation}

The chemical potentials of octet baryons $(b)$ with spin-1/2 and $\Delta$-resonances $(d)$ with spin- 3/2 are given by:
\begin{align}
    \mu_b =& \; \sqrt{p_{F_b}^2 + m_b^{*^2}} + g_{\omega b}\omega_0 + g_{\rho b}\boldsymbol{\tau}_{b3}\rho_{03} + \nonumber \\& g_{\phi b}\phi_0 + \Sigma^r \\
    \mu_d =& \; \sqrt{p_{F_d}^2 + m_d^{*^2}} + g_{\omega d}\omega_0 + g_{\rho d}\boldsymbol{\tau}_{d3}\rho_{03} + \Sigma^r
\end{align}
$\Sigma^r$ represents the self-energy re-arrangement term and is given by :
\begin{align}
    \Sigma^r =& \; \sum_b\bigg[\pdv{g_{\omega b}}{n}\omega_0 n_b - \pdv{g_{\sigma b}}{n}\sigma n^s_b + \pdv{g_{\rho b}}{n}\rho_{03}\boldsymbol{\tau}_{b3}n_b \nonumber \\&+ \pdv{g_{\phi b}}{n}\phi_0 n_b \bigg] + \sum_d \bigg[\pdv{g_{\omega d}}{n}\omega_0 n_d -\pdv{g_{\sigma d}}{n}\sigma n^s_d \nonumber \\& + \pdv{g_{\rho d}}{n}\rho_{03}\boldsymbol{\tau}_{d3}n_d \bigg]
\end{align}
where $n= \sum_b n_b + \sum_d n_d $ is the total vector (baryon) number density. $\Sigma^r$ is required in case of density dependent coupling models in order to maintain thermodynamic consistency \cite{PhysRevC.64.055805}. The chemical potential of s-wave condensates of (anti)kaons is given by:
\begin{align}
    \mu_{K^-} =& \; \sqrt{m_L^{*^2}+ |q_{K^-}|B} - g_{\omega K}\omega_0 - \frac{1}{2}g_{\rho K}\rho_{03} + g_{\phi K} \phi_0 \\
    \mu_{\bar{K}^0} =& \; m^*_K - g_{\omega K}\omega_0 + \frac{1}{2}g_{\rho K}\rho_{03} + g_{\phi K} \phi_0 
\end{align}
Threshold condition for the onset of the $i^{th}$ baryon is given by:
\begin{equation}
    \mu_i =\; \mu_n - q_i \mu_e
\end{equation}
with $\mu_e= \mu_n - \mu_p$ being the electron chemical potential. $q_i$ refers to the charge of the $i^{th}$ baryon.

The threshold condition for the appearance of (anti)kaons is given by:
\begin{align}
    \mu_{K^-} =& \; \mu_e = \; \mu_n- \mu_p \\ 
    \mu_{\bar{K}^0} =& \; 0
\end{align}
where $\mu_{K^-}$ and $\mu_{\bar{K}^0}$ are the chemical potentials of $K^-$ and $\bar{K}^0$, respectively. Muons $(\mu^-)$ appear when the chemical potential of electrons reaches the rest mass of muons $[\mu_e= m_\mu ]$. 

The matter inside NS is electrically neutral, with the charge neutrality condition gven by:
\begin{equation}
    \sum_b q_b n_b + \sum_d q_d n_d -n_e - n_\mu - n_{K^-} = 0
\end{equation}
The total energy density of the nuclear matter is given by:
\begin{align}
    \varepsilon =& \; \sum_b\varepsilon_b + \sum_d\varepsilon_d + \sum_l\varepsilon_l + \frac{1}{2} m_\sigma^2 \sigma^2 + \frac{1}{2}m_\omega^2 \omega_0^2 \nonumber \\& + \frac{1}{2}m_\rho^2\rho_{03}^2 + \frac{1}{2}m_\phi \phi_0^2 + \varepsilon_{\bar{K}}
\end{align}
where $\varepsilon_{\bar{K}}$ is the kaonic contribution to the total energy density and is given by:
\begin{equation}
    \varepsilon_{\bar{K}}= \;m^*_K ( n_{K^-} + n_{\bar{K}^0})
\end{equation}
From the Gibbs-Duhem relation, we get the matter pressure as:
\begin{equation}
    P = \; \sum_b\mu_b n_b + \sum_d \mu_d n_d + \sum_l \mu_l n_l - \varepsilon
\end{equation}
(Anti)kaons, being s-wave condensates, do not contribute explicitly to the matter pressure. $\Sigma^r$ contributes explicitly only to the matter pressure.

\subsection{Star structure}

\begin{table*}
\centering
\caption{Table for nuclear saturation properties for the three parametrizations- DD-ME2, DD-2 and DD-MEX. \label{table 1}}
\begin{tabular}[t]{cccccccccccccccccccccccccccc} 
\toprule \toprule
Parametrization  &&&$n_0$ &&&$E/A$ &&&$K_0$ &&&$E_{sym}$ &&&$L_{sym}$ &&&$m^*_N/m_N$ &&&$m_{\sigma}$ &&&$m_N$ \\
 &&&$(fm^{-3})$ &&&$(MeV)$ &&&$(MeV)$ &&&$(MeV)$ &&& $(MeV)$ &&& &&&$(MeV)$ &&&$(MeV)$ \\
\midrule
DD-ME2 &&&0.152    &&&-16.14  &&&250.89  &&&32.30  &&&51.253 &&&0.572  &&&550.124  &&&938.90 \\
DD-2   &&&0.149065 &&&-16.02  &&&242.70  &&&32.73  &&&54.966 &&&0.5625 &&&546.2124 &&&939.56\\
DD-MEX &&&0.152    &&&-16.097 &&&267.059 &&&32.269 &&&49.576 &&&0.556  &&&547.3327 &&&939.00\\
\bottomrule \bottomrule
\end{tabular}
\end{table*}

\begin{table*}
\centering
\caption{Table for parameter values for DD-ME2, DD-2 and DD-MEX parametrizations. The masses of $\omega$, $\rho$ and $\phi$ mesons are 783 MeV, 763 MeV and 1019.45 MeV, respectively, and they are same for all the three parametizations. \label{table 2}}
\begin{tabular}[t]{ccccccccccccccccccccccccc} 
\toprule \toprule
Parametrization &&&Meson($i)$ &&&$a_i$ &&&$b_i$ &&&$c_i$ &&&$d_i$ &&&$g_{iN}$\\
\midrule
  &&&$\sigma$   &&&1.3881  &&&1.0943  &&&1.7057  &&&0.4421  &&&10.5396\\
DD-ME2 &&&$\omega$   &&&1.3892  &&&0.9240  &&&1.4620  &&&0.4775  &&&13.0189\\
        &&&$\rho$     &&&0.5647  &&&&&&&&&                        &&&7.3672\\
\\
    &&&$\sigma$   &&&1.3576  &&&0.6344  &&&1.0053  &&&0.5758  &&&10.6866\\
DD-2        &&&$\omega$   &&&1.3697  &&&0.4964  &&&0.8177  &&&0.6384  &&&13.3423\\
        &&&$\rho$     &&&0.5189  &&&&&&&&&                        &&&7.2538\\       
\\
  &&&$\sigma$   &&&1.3970  &&&1.3349  &&&2.0671  &&&0.4016  &&&10.7067\\
DD-MEX        &&&$\omega$   &&&1.3926  &&&1.0919  &&&1.6059  &&&0.4556  &&&13.3388\\
        &&&$\rho$     &&&0.6202  &&&&&&&&&                        &&&7.2380\\
\bottomrule \bottomrule
\end{tabular}
\end{table*}

The solution for the Einstein's equations for general relativity for a static and spherically symmetric star gives us the Tolman-Oppenheimer-Volkoff (TOV) equations. These equations are then numerically solved for a particular EoS to obtain the mass-radius relationship of the NS. The TOV equations are as follows \cite{glendenning2012compact}:

\begin{align}\label{eqn.33}
\begin{split}
    \frac{dP(r)}{dr} &= -\frac{[P(r) + \varepsilon(r)][M(r) +4\pi r^3 P(r)]}{r[r-2M(r)]} \\
    \frac{dM(r)}{dr} &= 4\pi r^2 \varepsilon (r)
\end{split}    
\end{align}

where $M(r)$ is the gravitational mass included within radius $r$. The TOV equations are solved with the boundary conditions $M(0)=0$ and $P(R)=0$, where $R$ is the radius of the NS.
The presence of strong magnetic field, however, distorts the spherical symmetry of the star structure. The most general coupled set of equations determining the spherically symmetric star structure, as derived by Bowers and Liang, is given as \cite{1974ApJ...188..657B}:
\begin{align} \label{eqn.34}
    \frac{dM(r)}{dr} &= 4\pi r^2 \varepsilon \nonumber \\
    \frac{d\Phi(r)}{dr} &= \left(1-\frac{2M}{r}\right)^{-1}\left(\frac{M}{r^2} + 4\pi  P_r r\right) \nonumber \\
    \frac{dP_r(r)}{dr} &= -(\varepsilon + P_r)\frac{d\Phi}{r} + \frac{2}{r}(P_{\perp} - P_r)
\end{align}
where $P_r$ and $P_{\perp}$ are the radial and tangential pressure components, respectively, and  $\Phi$ is the Newtonian gravitational potential at the Newtonian limit. The most general energy-momentum tensor, considering spherical symmetry, is :
\begin{equation}\label{eqn.35}
    T^{\mu \nu} = diag( \varepsilon, P_r, P_{\perp}, P_{\perp})
\end{equation}
However, following the same argument given in \cite{PhysRevC.99.055811}, $T^{\theta \theta} \neq T^{\phi \phi}$ (Eqns.(23d) $\&$ (23e) in \cite{10.1093/mnras/stu2706}) for the case of the electromagnetic energy-momentum tensor. This is in contradiction to the spherical symmetry assumption in Eqn.(\ref{eqn.35}). Also, the last term in Eqn.(\ref{eqn.34}) diverges at the origin since $\lim_{r\to 0} (T^{rr} - T^{\theta \theta}) \neq 0$. Therefore, we see that spherically symmetric solutions cannot exactly describe the star structure in the presence of magnetic field. Although Eqn.(\ref{eqn.33}) provides a good approximation of the mass-radius relation of magnetized NS \cite{PhysRevC.99.055811}, for central magnetic fields close to $10^{18}$ G the deformity becomes too large for Eqn.(\ref{eqn.33}) to be used \cite{Dexheimer_2017}. Thus, we refrain from using the spherically symmetric TOV equations for magnetic field strengths of $> 10^{17}$ G.

\section{Results and discussion}\label{Sec 3}
\subsection{Parametrizations}

In this DD-RMF model, the density dependant nature of the meson-nucleon coupling constants for $\sigma$ and $\omega$ mesons is given by:
\begin{align}
    g_{iN}(n) = \;g_{iN}(n_0)f_i(x)\;, \;\;\;\;\;  i=\sigma,\omega
\end{align}
where $n$ and $n_0$ are the total baryon number density and the nuclear saturation density, respectively. $N$ refers to nucleons. The variable $x = n/n_0$. $f_i(x)$ is defined as:
\begin{equation}\label{36}
    f_i(x)= a_i \frac{1 +b_i(x+d_i)^2}{1 + c_i(x+d_i)^2}.
\end{equation}
The meson-nucleon coupling constant for $\rho$ meson is given by:
\begin{equation}\label{37}
    g_{\rho N}=\; g_{\rho N}(n_0)e^{-a_\rho(x-1)}.
\end{equation}
The $\phi$ meson does not couple with nucleons and, thus, $g_{\phi N}=0$. We take $m_\omega=783$, $m_\rho=763$ and $m_{\phi} = 1019.45$ MeV. 

For the calculation of the scalar meson-hyperon coupling constants, we consider the optical potentials of $\Lambda$, $\Xi$ and $\Sigma$ to be $U_{\Lambda}=-30 \; MeV$, $U_{\Xi}=-14 \; MeV$ and $U_{\Sigma}=+30 \;MeV$, respectively \cite{particles3040043}. $\Sigma$ hyperons, having a repulsive optical potential, do not appear in the range of densities considered in our work. 

In case of the vector meson-hyperon density dependent vector coupling constants, we employ SU(6) \cite{PhysRevC.64.025804} symmetry and get the following relations:
\begin{align}
\begin{split}
    &\frac{1}{2}g_{\omega\Lambda} =g_{\omega\Xi}= \frac{1}{2}g_{\omega\Sigma}=\frac{1}{3}g_{\omega N}\\
    &2g_{\phi\Sigma}=2g_{\phi\Lambda}=g_{\phi\Xi}=-\frac{2\sqrt{2}}{3}g_{\omega N}\\
    &\frac{1}{2}g_{\rho\Sigma}=g_{\rho\Xi}=g_{\rho N}\\
    &g_{\rho\Lambda}=0
\end{split}    
\end{align}

For the scalar meson-$\Delta$ coupling constants, we fix the $\Delta$-potential to $V_\Delta= \frac{4}{3}V_N$, which gives $R_{\sigma\Delta} = g_{\sigma \Delta} / g_{\sigma N}= 1.16$. $V_N$ stands for the nucleon potential. For $\Delta-$resonances, the vector coupling constants are given by \cite{PhysRevC.100.015809}:
\begin{align}
    g_{\omega\Delta}=1.1g_{\omega N}, \;\;\;g_{\rho\Delta}=g_{\rho N}.
\end{align}
$\phi$ meson does not couple with $\Delta-$resonances and, thus, $g_{\phi\Delta}=0$.

The calculation for the scalar meson-(anti)kaon coupling constants is explained in \cite{PhysRevD.102.123007}. Several works \cite{WAAS1997287,PhysRevC.60.024314,KOCH19947,LUTZ199812,SCHAFFNERBIELICH2000153} have provided the $K^-$ optical potential ($U_{\bar{K}}$) in the range $-200 \leq U_{\bar{K}} \leq -40$ MeV. In this work, we have chosen $U_{\bar{K}}=-130$ MeV. The determination of  vector meson-(anti)kaon coulping constants is given in \cite{PhysRevC.90.015801,PhysRevC.87.045802}. They are density independent and are given by the relations:
\begin{align}
    g_{\omega K}=\frac{1}{3}g_{\omega N}, \;\;\;g_{\rho K}=g_{\rho N},\;\;\;g_{\phi K}=4.27
\end{align}

\textcolor{black}{It is to be noted that the more general SU(3) symmetry has also been implemented in many past works \cite{PhysRevC.88.015802, 2012PhRvC..85f5802W, 2020PhRvD.102f3008M, 1994nucl.th...1004R}, in-lieu of SU(6), to determine the hyperon-vector meson coupling parameters.
Incorporating SU(3) symmetry brings into picture free parameters with uncertainties.
%However, we have decided to neglect SU(3) symmetry because coupling constants determined using SU(3) symmetry were originally used to increase the maximum mass in order to satisfy the observational constraints.
In our current model, \textcolor{black}{however},we are successful in satisfying the observational constraints with SU(6) symmetry. Another thing to note is that coupling constants determined using SU(3) not only increase the maximum mass of NSs but also increase their radius, which in turn makes the stars violate the tidal deformability constraints. So, we proceed with SU(6) symmetry in this paper.} 

In this work, we use three different density dependent parametrizations:  DD-ME2, DD2 and DD-MEX. The three parametrizations are framed to reproduce the nuclear matter properties at $n_0$. The nuclear saturation properties as well as the masses of nucleons and $\sigma$ mesons for the three parametrizations are shown in Table \ref{table 1}. In the table, $E/A$, $K_0$, $E_{sym}$, $L_{sym}$, $m_N$, $m^*_N$ and $m_{\sigma}$ stand for binding energy per nucleon, compression modulus, symmetry energy coefficient, slope parameter of $E_{sym}$, mass of nucleons, effective mass of nucleons and mass of $\sigma$ mesons, respectively. All the properties in Table \ref{table 1} are evaluated at nuclear saturation density ($n_0$). The values of the coefficients in Eqns.\eqref{36} and \eqref{37} for the parametrizations DD-ME2 \cite{PhysRevC.71.024312} , DD2 \cite{PhysRevC.81.015803} and DD-MEX \cite{TANINAH2020135065} are given in Table \ref{table 2}.
\textcolor{black}{$\bar{K}$ condensates can appear via both first order and second order phase transitions depending upon the (anti)kaon optical potential in nuclear symmetric matter \cite{glendenning2012compact}.}
However, we note that with the discussed parametrizations only the second order phase transition occurs \cite{PhysRevD.102.123007}.

\begin{figure}[t!]
    \includegraphics[width=8.5cm, keepaspectratio]{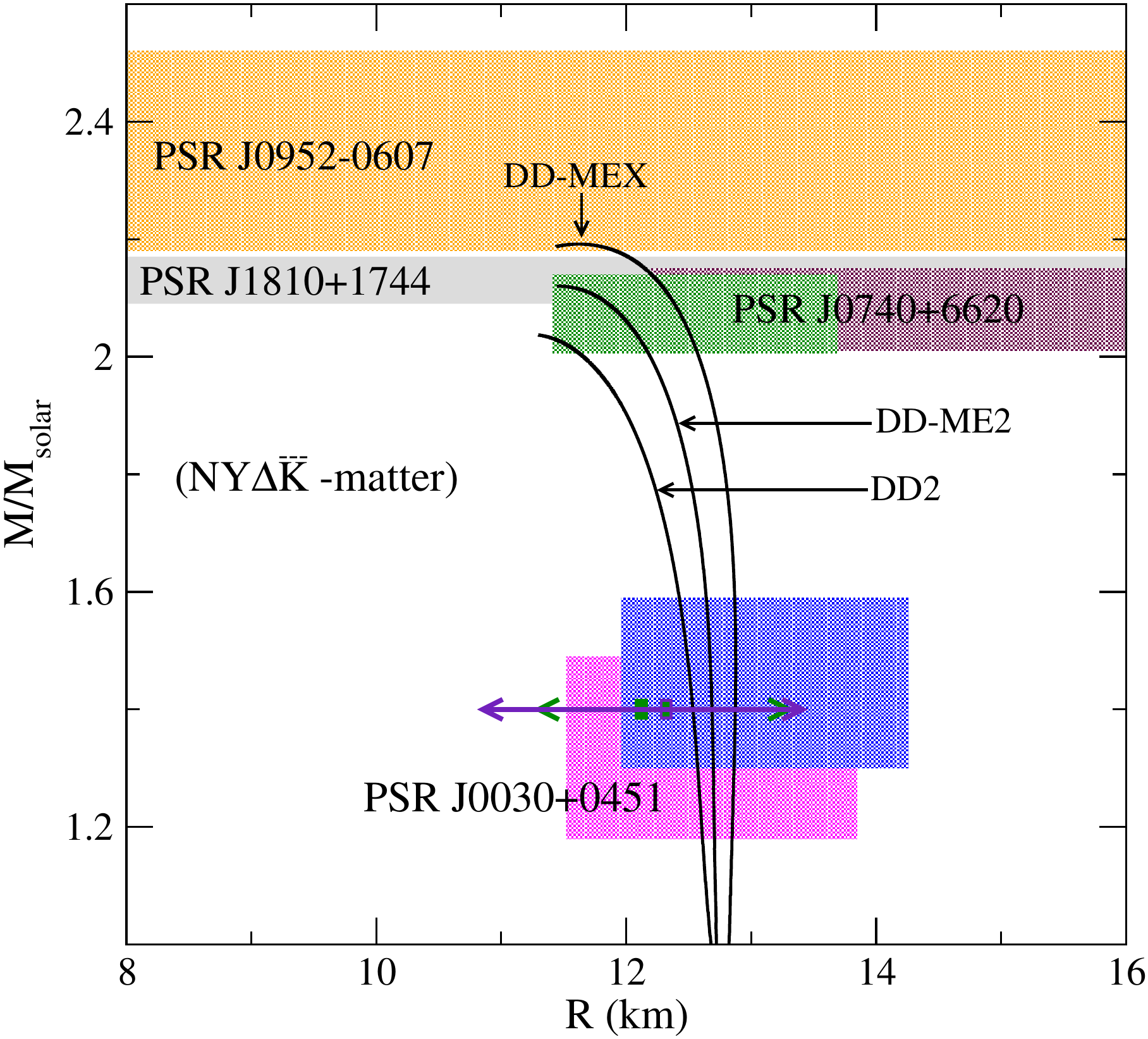}
    \captionof{figure}{Mass-radius relationship of NS for the matter composition $N\bar{K}Y\Delta$ for the parametrizations DD-ME2, DD-2 and DD-MEX. The shaded regions illustrate the observational constraints from PSR J0740+6620 \cite{2021ApJ...918L..27R, 2021ApJ...918L..28M}, PSR J1810+1744 \cite{2021ApJ...908L..46R}, PSR J0030+0451 \cite{Miller_2019, Riley_2019} and PSR J0952-0607 \cite{2022ApJ...934L..17R}. \textcolor{black}{The joint radius constraints from PSR J$0030+0451$ and the GW170817 event data for a typical 1.4 M$_\odot$ NS are represented by the horizontal lines \cite{Jiang_2020, 2020PhRvD.101l3007L}}.}
    \label{fig.1}
\end{figure}

With these three parametrizations we note the star structure, ignoring the effect of magnetic field, from the mass-radius relation as shown in Fig. \ref{fig.1}. From the figure, it is evident that the NSs composed of matter including hyperons, $\Delta$-resonances and (anti)kaon condensates satisfy the so far obtained astrophysical constraints on mass-radius relation for the DD-MEX and DD-ME2 parametrizations. Even though it does not satisfy the most recent observation PSR J0952-0607, we still keep the DD-ME2 parametrization because it satisfies all the other observational constraints. Along with this, these two parametrizations also obey the maximum limit of mutual tidal deformability obtained from gravitational wave observations of binary NS merger event GW170817, which is evident from the Fig. \ref{fig.2}. So, in our present study of the effect of magnetic field on the NS composed of matter containing hyperons, $\Delta-$resonances and (anti)kaon condensates, we choose these two parametrizations which are compatible with astrophysical observations.

\begin{figure}
    \includegraphics[width=8.5cm, keepaspectratio]{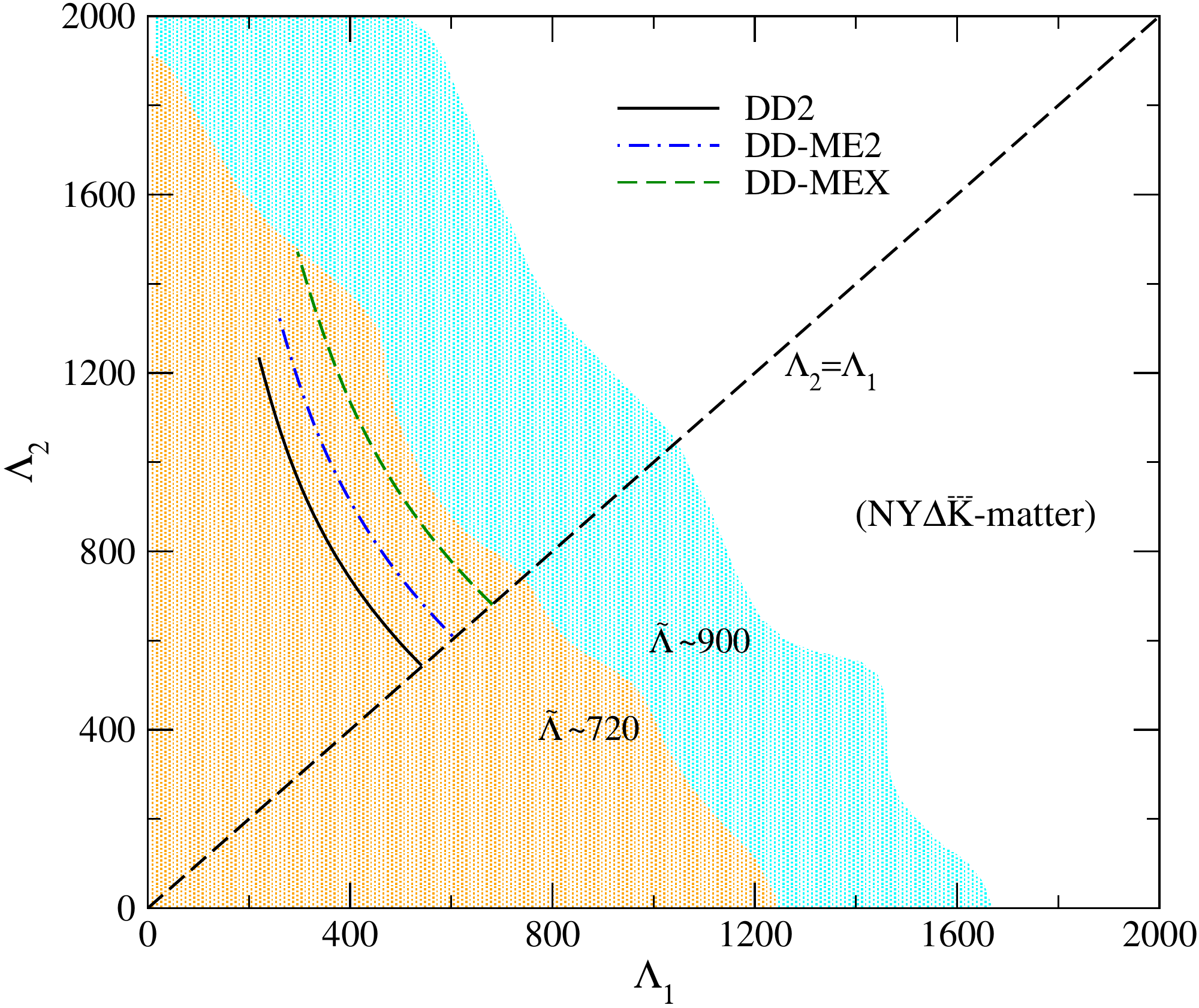}
    \caption{Plot for the tidal deformablities $\Lambda_1$ and $\Lambda_2$ for the matter composition $N\bar{K}Y\Delta$ and considering a fixed chirp mass, $\mathcal{M}= 1.188 M_\odot$. The $\Lambda_1$ and $\Lambda_2$ correspond to the stars of masses $M_1$ and $M_2$, respectively, of the binary system observed in GW170817 event. The shaded regions represent $\Tilde{\Lambda} \sim 900$(TaylorF2) and $\Tilde{\Lambda} \sim 720$(PhenomPNRT) upper bounds at 90\% confidence level \cite{2017PhRvL.119p1101A, PhysRevX.9.011001}. \label{fig.2} }
    \label{fig.2}
\end{figure}

\textcolor{black}{It is to be noted that recent studies on the correlation between the slope of symmetry energy and the neutron skin thickness of \ce{^{208}Pb}, obtained from the combined results of PREX and PREX-II experiments,} \textcolor{black}{indicate the range of $L_{sym}= (106 \pm 37)$ MeV $\&$ $E_{sym}= (38.1 \pm 4.7)$ MeV \cite{PhysRevLett.126.172503}. The corresponding iso-vector coupling values for the PREX-II estimations of symmetry energy are adapted from Ref.-\cite{2022arXiv220302272B}. However, for these} \textcolor{black}{increased values of $L_{sym}$, the maximum attainable mass gets reduced and tends to fall short of the recent observational constraints. Considering the symmetry energy coefficient values as $L_{sym}= 70$  MeV and $E_{sym}=38.1$ MeV, we get a maximum mass of 2.156 $M_{\odot}$ (as opposed to our originally calculated value of 2.192 $M_{\odot}$) and 2.088 $M_{\odot}$ (as opposed to our originally calculated value of 2.120 $M_{\odot}$) for DD-ME2 and DD-MEX parametrizations, respectively. Varying the symmetry energy parameter within the above provided range further decreases the maximum mass. From nuclear physics experiment also, the range of $L_{sym}$ and $E_{sym}$ obtained from these experiments are not very reliable as another recent experiment to determine the neutron skin thickness in $^{48}$Ca-isotope (CREX) \cite{PhysRevLett.129.042501} reports the same to be much smaller and in disagreement with the PREX estimations. Due to the violation of observational constraints for these values of $L_{sym}$ and $E_{sym}$, we refrain from proceeding with them in further discussion in this paper.}

\subsection{Magnetic Field Profile}
    We model the magnetic field inside NS by adopting a magnetic field profile which is consistent with the Einstein-Maxwell field equations. One such magnetic field profile has been obtained by finding the solutions of the Einstein-Maxwell field equations with magnetostatic equilibrium for EoS from several nuclear models and then taking a polynomial fit of the monopolar part of the norm of the magnetic field profiles obtained. This is the universal profile given by \cite{PhysRevC.99.055811}
\begin{equation}\label{eqn.41}
    B(x)=B_m(1-1.6x^2-x^4+4.2x^6-2.4x^8)
\end{equation}
where $x\equiv r/r_{mean}$, $r$ being the radial distance, $r_{mean}$ is the mean radius of the star and $B_m$ is the field strength at the centre of the star. 
%This \textcolor{black}{profile has been} obtained \textcolor{black}{by}, as shown in \cite{PhysRevC.99.055811}, by solving the Einstein-Maxwell equations together with magnetostatic equilibrium for EoS from several nuclear models and than taking a polynomial fit of the obtained profiles. 
This profile, however, is for a star with an approximate monopolar magnetic structure and does not incorporate the dipolar structure. 
%as Eqn. \ref{eqn.40} does. 

Another such profile, obtained by taking a quadratic fit of the solutions of Einstein-Maxwell field equations assuming a poloidal magnetic field for EoS from three different nuclear models and two different values of magnetic dipole moment, is given by \cite{DEXHEIMER2017487, 2021ApJ...917...46R}:
\begin{equation}\label{eqn.40}
    B(\mu_B)= \; \frac{(a + b\mu_B + c\mu_B^2)}{B_c}\mu
\end{equation}
where $\mu_B$ is the baryon chemical potential and $\mu$ is the dipole magnetic moment of the NS. $\mu_B$ and $\mu$ are in units of $MeV$ and $Am^2$, respectively, to get $B(\mu_B)$ in units of gauss ($G$). $B_c = \; 4.414 \times 10^{13}$ G is the critical field of electron. The values of the coefficients $a$, $b$, $c$ for a star of mass $2.2 M_{\odot}$ are as follows:
\begin{align}
    &a=\; -0.769 \; G^2/(Am)^2 \nonumber \\&
    b= \; 1.2\times 10^{-3}\; G^2/(Am^2\;MeV)\;, \nonumber \\&
    c= \; -3.46 \times 10^{-7} \; G^2/(Am^2 \; MeV^2) \nonumber
\end{align}
%eq. \eqref{eqn.40} was obtained, as shown in \cite{DEXHEIMER2017487}, by solving the Einstein-Maxwell equations assuming a poloidal magnetic field for EoS from three different nuclear models and two different values of magnetic dipole moment, and then taking a quadratic fit of the obtained magnetic field profiles.
\textcolor{black}{Both the magnetic field profiles are derived using input from several different EoSs \cite{PhysRevC.99.055811, DEXHEIMER2017487}. However, none of the EoSs used are for the entire range of particles considered in our present work. Even so, we believe that the variety of EoSs used, a few of which are very close in matter composition to our present work, in deriving the field profiles make Eqns.\eqref{eqn.41} and \eqref{eqn.40} viable candidates for use in our present nuclear model in a self-consistent manner. To be completely accurate in maintaining self-consistency of the field profile with the nuclear model, our EoS needs to be used as input in deriving the deriving field profiles, which is beyond the scope of our present work.}

\begin{figure}[h!]
    \includegraphics[width=8.5cm, keepaspectratio]{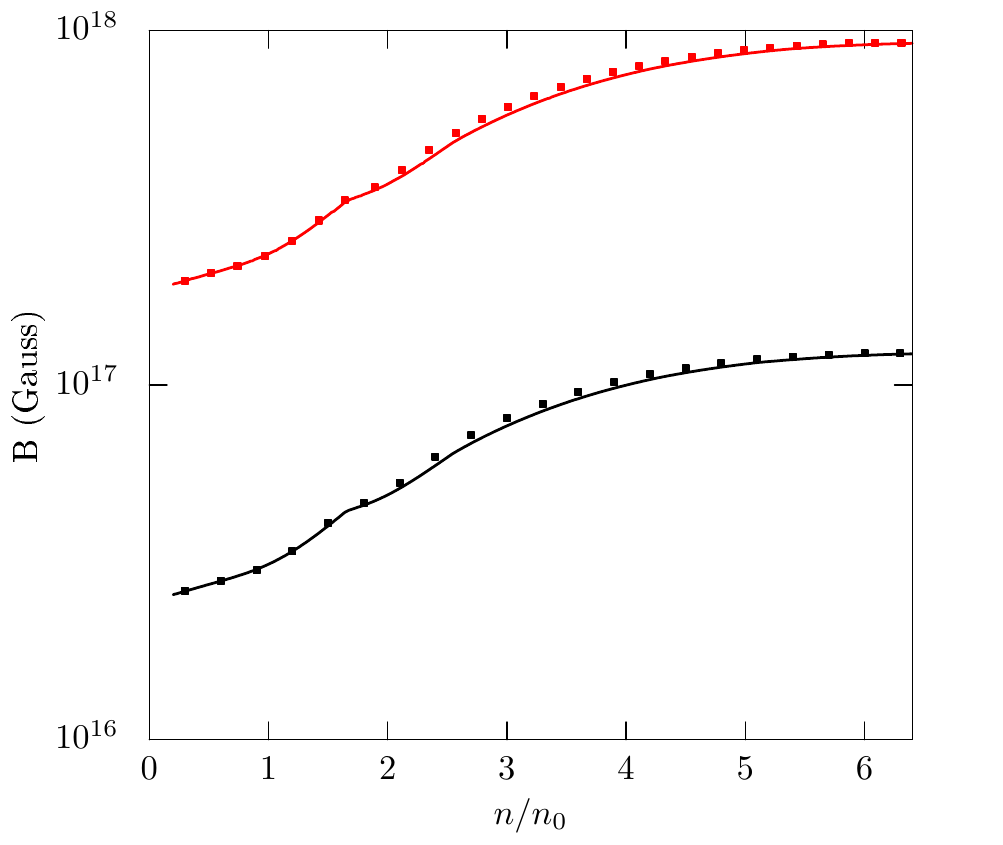}
    \captionof{figure}{The variation of magnetic field with the normalized number density $n/n_0$ corresponding to the magnetic field profile given by Eqn. \eqref{eqn.40}. The upper curves are for $\mu = 1.5\times 10^{32} \; Am^2$ and the lower curves are for $\mu = 2\times 10^{31} \; Am^2$. The solid lines represent DD-ME2 parametrization while the dotted lines represent DD-MEX parametrization.%Magnetic field profiles, $B$, in the polar direction \textcolor{black}{ as functions of total number density normalized to nuclear saturation density . The curves for the two dipole magnetic moments, $\mu =  \; Am^2$ and ,  }
    }
    \label{fig.3}
\end{figure}

\textcolor{black}{We note that Eqn.\eqref{eqn.40},} being a function of baryon chemical potential, also avoids discontinuities in the field during phase transitions. Hence, we choose this profile in our following calculations. Here, we consider two values of the dipole magnetic moment, $\mu= 2\times 10^{31} Am^2$ and $\mu= 1.5\times 10^{32} Am^2$, which give central magnetic fields of around $1.2{\times}10^{17}$ and $0.9{\times}10^{18} G$, respectively, and surface magnetic fields of around $2.5{\times}10^{16} G$ and $2{\times}10^{17} G$, respectively. The magnetic field profiles for each case are shown in Fig. \ref{fig.3}. 

\begin{figure*}
   \centering
\includegraphics[width=16.5cm,keepaspectratio ]{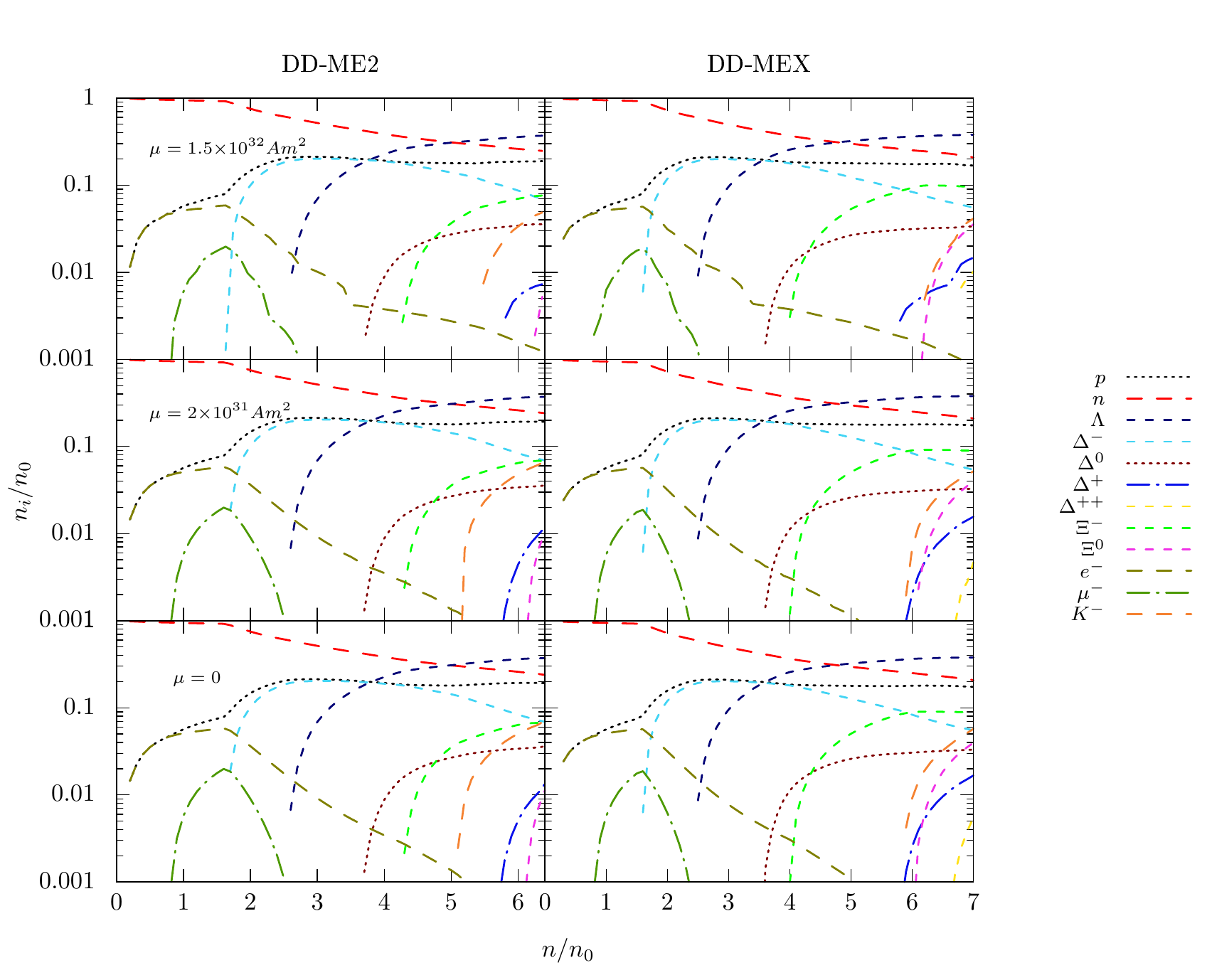}
    \caption{Variation of particle number density normalized to the nuclear saturation density $(n_i/n_0)$ with the total number density normalized to $n_0$ $(n/n_0)$. Top panel: in the presence of magnetic field with $\mu= 1.5{\times}10^{32} Am^2$, middle panel: in the presence of magnetic field with $\mu= 2{\times}10^{31} Am^2$ and bottom panel: in the absence of magnetic field. Left panel: for DD-ME2 and right panel: DD-MEX parametrization.}
    \label{fig.4}
\end{figure*}

\begin{table*}
\centering
\caption{Threshold densities of various particles. The particle $\mu^-$ extinguishes at 2.77 and 2.59 for DD-ME2 and DD-MEX, respectively, for B field with $\mu=2{\times}10^{31}Am^2$, and the particle $\mu^-$ extinguishes at 2.79 and 2.62 for DD-ME2 and DD-MEX, respectively, for B field with $\mu=1.5{\times}10^{32}Am^2$ . Without magnetic field, the particle $\mu^-$ extinguishes at 2.77 and 2.58 for DD-ME2 and DD-MEX, respectively. \label{table 3} }
\begin{tabular}[t]{cccccccccccccccccccccccc} 
\toprule \toprule
&&& & \multicolumn{4}{c}{$\mu=1.5{\times}10^{32}Am^2$} 
&&& & \multicolumn{4}{c}{$\mu=2{\times}10^{31}Am^2$} 
&&& & \multicolumn{4}{c}{Without B field} \\
\cmidrule{5-8}\cmidrule{12-15}\cmidrule{19-22}
&&& &DD-ME2 && &DD-MEX &&& &DD-ME2 && &DD-MEX &&& &DD-ME2 && &DD-MEX\\
\midrule
$\mu^-$  &&& &0.75 && &0.75 
&&& &0.75 && &0.75 
&&& &0.75 && &0.75 \\
$\Lambda$  &&& &2.53
&& &2.42 
&&& &2.53
&& &2.42 
&&& &2.53 && &2.42 \\
$\Delta^-$  &&& &1.63 && &1.58
&&& &1.63 && &1.58
&&& &1.63 && &1.58 \\
$\Delta^0$  &&& &3.61 && &3.50 
&&& &3.61 && &3.50 
&&& &3.61 && &3.50 \\
$\bar{K}^-$  &&& &5.37 && &6.11 
&&& &5.11 && &5.87 
&&& &5.07 && &5.82 \\
$\Delta^+$  &&& &5.75 && &5.72 
&&& &5.64 && &5.71 
&&& &5.62 && &5.70 \\
$\Delta^{++}$  &&& &     && &6.70
&&& &     && &6.57
&&& &    && &6.55 \\
$\Xi-$  &&& &4.21 && &3.94 
&&& &4.24 && &3.97 
&&& &4.24 && &3.97 \\
$\Xi^0$  &&& &6.16 && &6.08 
&&& &6.08 && &6.00 
&&& &6.06 && &5.99 \\
\bottomrule \bottomrule
\end{tabular}
\end{table*}

\begin{figure*}
    \centering
    \includegraphics[width=15.5cm, keepaspectratio]{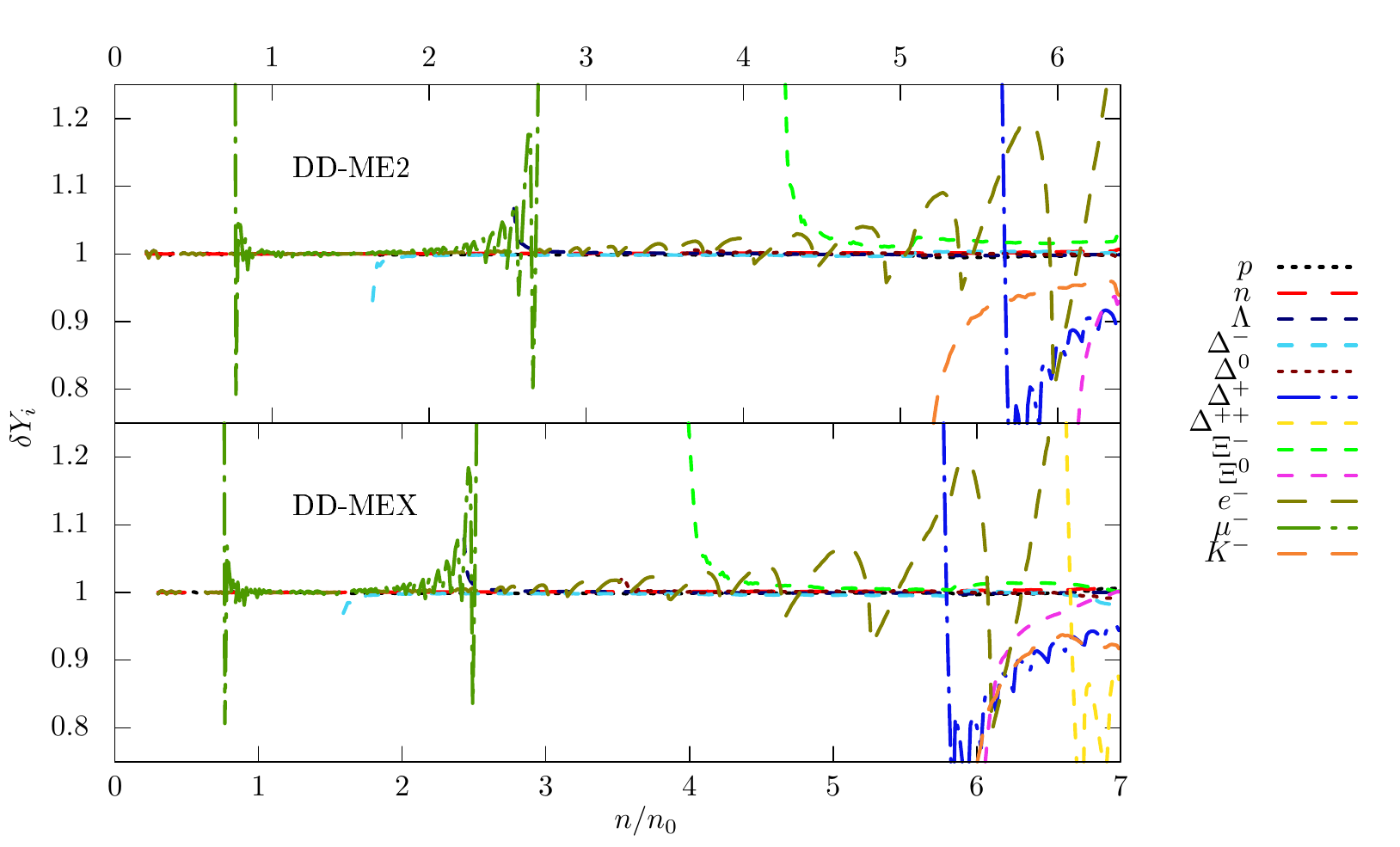}
    \caption{Variation of $\delta Y_i = n_i(B)/n_i(0)$ with the total number density normalized to $n_0$ ($n_i/n_0$) for the star dipole moment $\mu=2{\times}10^{31} Am^2$. Upper panel: DD-ME2 parametrization and lower panel: DD-MEX parametrization.}
    \label{fig.5}
\end{figure*}

\begin{figure*}
    \centering
    \includegraphics[width=15.5cm, keepaspectratio]{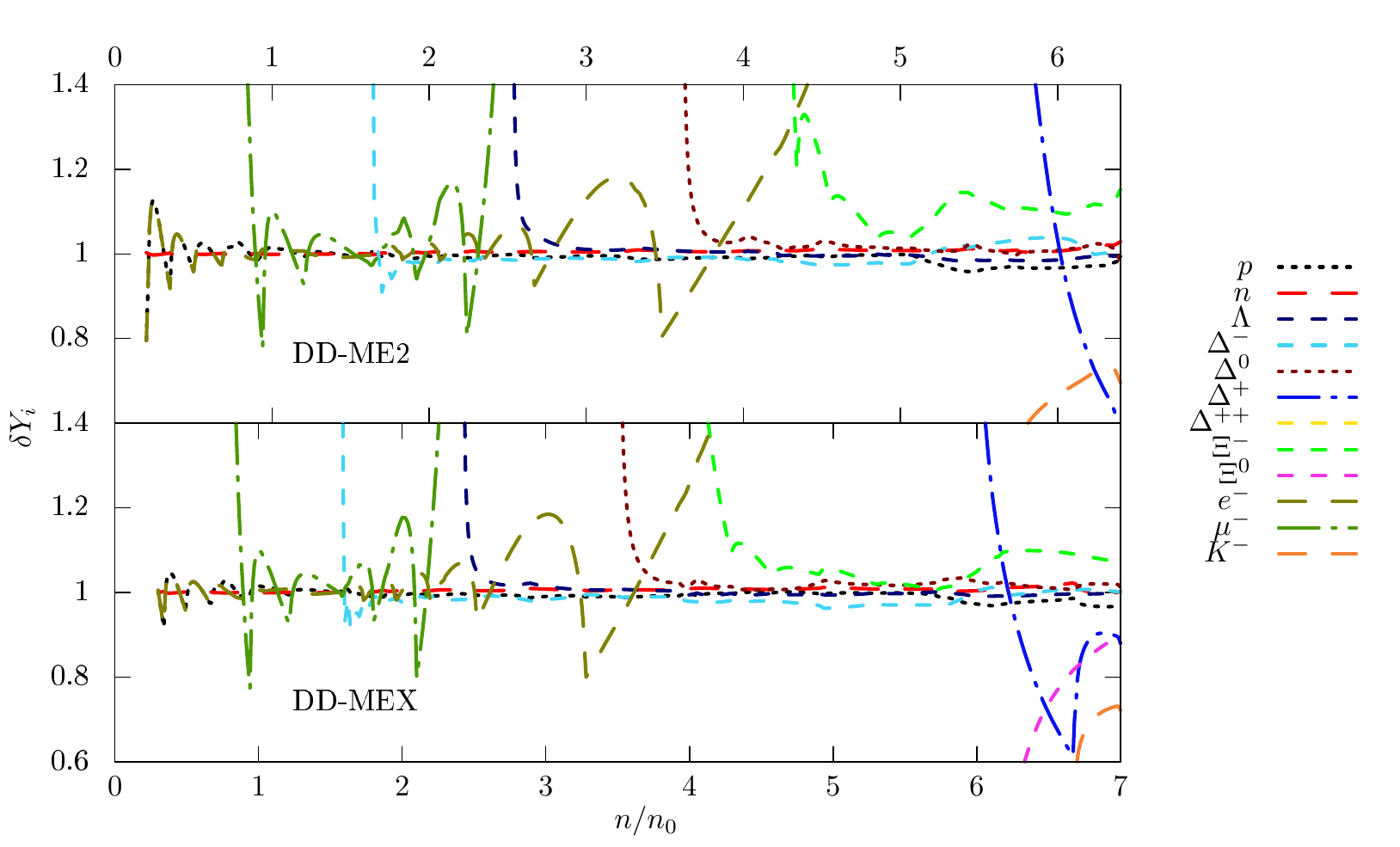}
    \caption{Variation of $\delta Y_i = n_i(B)/n_i(0)$ with the total number density normalized to $n_0$ ($n_i/n_0$) for the star dipole moment $\mu=1.5{\times}10^{32} Am^2$. Upper panel: DD-ME2 parametrization and lower panel: DD-MEX parametrization.}
    \label{fig.6}
\end{figure*}

\begin{figure*}
    \centering
    \includegraphics[width=17.5cm, keepaspectratio]{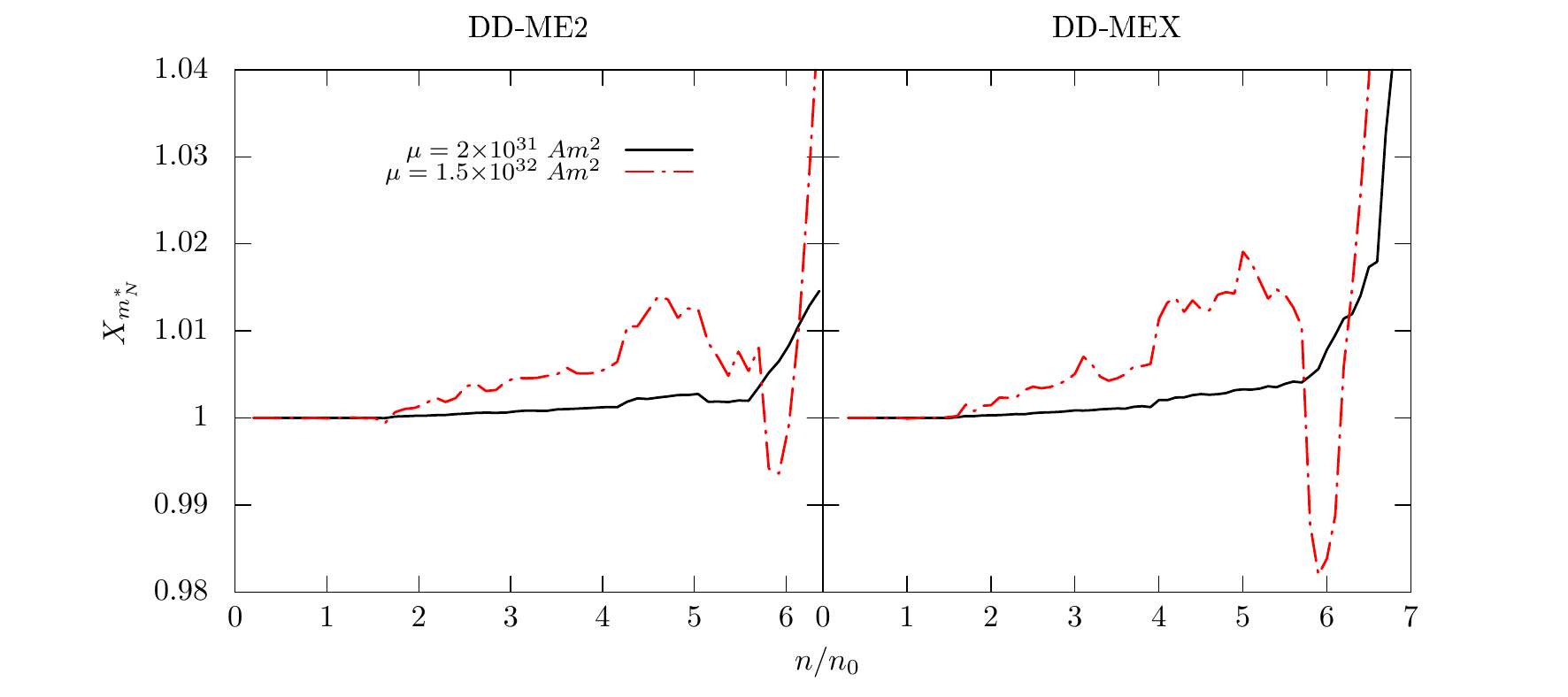}
    \caption{Plot for the ratio $X_{m^*_N} \equiv m^*_N(B)/m^*_N(0)$ of the effective Dirac mass of nucleons in the presence of magnetic field to it's value in the absence of magnetic field as function of total number density normalized to $n_0$ $(n/n_0)$ for the star dipole moments $\mu=2{\times}10^{31}Am^2$ and $\mu=2{\times}10^{31}Am^2$. The plots are given for the matter matter composition $N\bar{K}Y\Delta$ and for the two parametrizations: DD-ME2 (left panel) and DD-MEX (right panel). }
    \label{fig.7}
\end{figure*}

\subsection{Matter and star with magnetic field}

The particle fraction profiles for $NY\bar{K}\Delta$ are illustrated in Fig. \ref{fig.4} for the two parametrizations, both with and without magnetic field. The particle population at all densities satisfies the two conditions- charge neutrality and baryon number conservation. As can be inferred from the figure, at the initial densities, the charge neutrality is maintained by protons $(p)$ and leptons- electrons $(e^-)$ and muons $(\mu^-)$. The $e^-$ and $\mu^-$ populations clearly decreases from the onset of the negative $\Delta$-resonance ( $\Delta ^-$) till they eventually disappear or become insignificantly sparse. This is because $\Delta^-$ is energetically more favourable than the leptons and thus take their place in maintaining charge neutrality. The onset of $\Xi^-$ further contributes to the decrease in $e^-$ population. The $\Delta^-$ population, however, starts to decrease with the onset of $\Xi^-$ and decreases more heavily with the onset of $K^-$. At the extreme higher end of the density range, the negative charges are provided by $K^-$, $\Delta^-$ and $\Xi^-$ while the positive charges are provided by $p$, $\Delta^+$ and $\Delta^{++}$, in a way such that charge neutrality remains intact. However, we note that $\Delta^{++}$ only appears in the case for DD-MEX parametrization and is absent in the case of DD-ME2. 

\begin{figure*}
    \centering
    \advance\leftskip -1.4 cm
    \includegraphics[width=18.5cm, keepaspectratio]{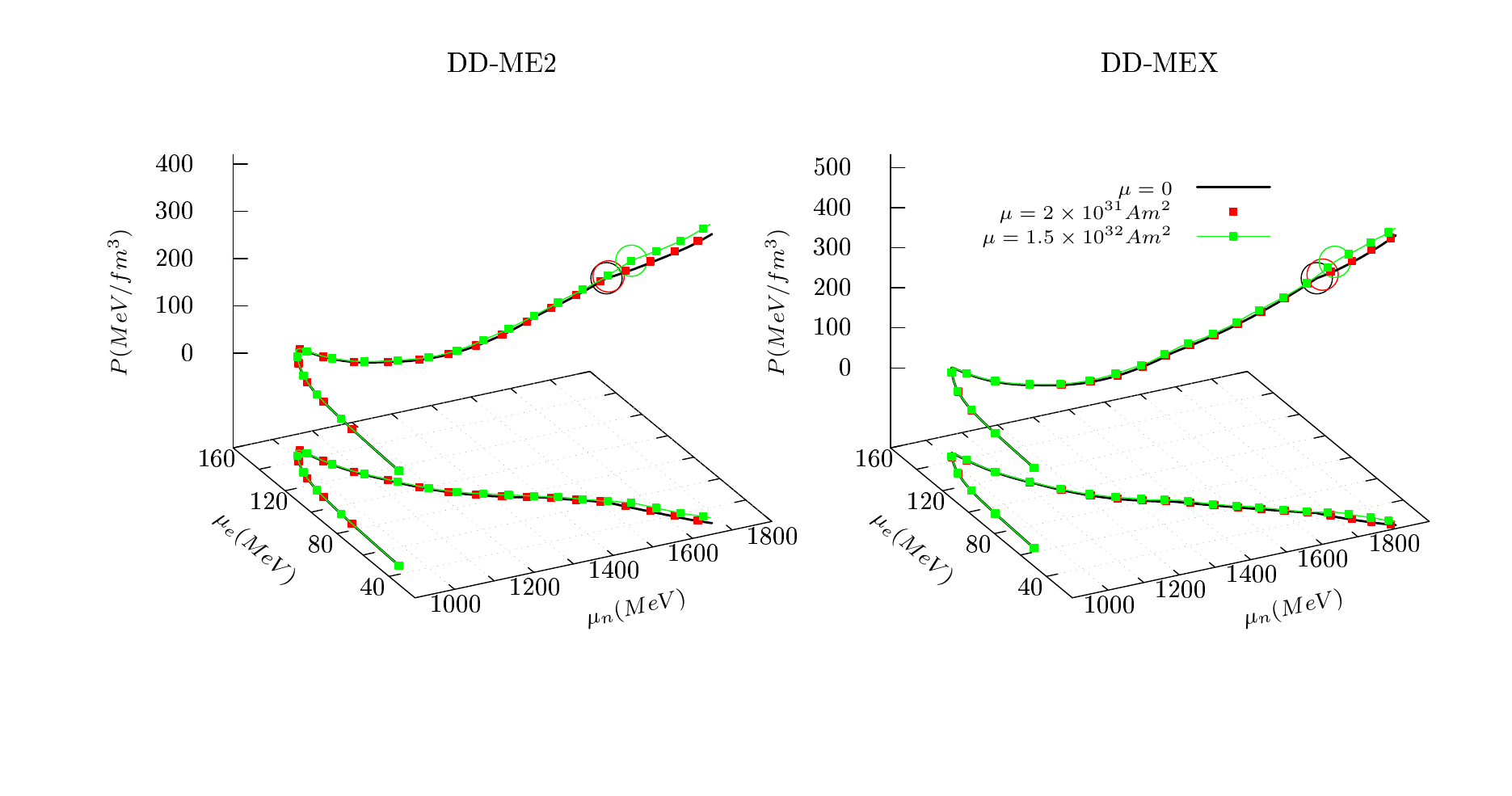}
    \caption{\textcolor{black}{3D plot of pressure density as a function of electron chemical potential ($\mu_e$) and neutron chemical potential ($\mu_n$), along with its projection on the xy-plane. The left and right panels are for DD-ME2 and DD-MEX parametrizations, respectively. The circles mark the onset of (anti)kaon condensates: black circle for $\mu=0$, red circle for $\mu=2{\times}10^{31}Am^2$ and green circle for $\mu=1.5{\times}10^{32}Am^2$ .}}
    \label{fig.10}
\end{figure*}

The effect of magnetic field on the particle population can be better appreciated by looking at Fig. \ref{fig.5} and Fig. \ref{fig.6}. They show the ratio $\delta Y_i \equiv n_i(B)/n_i(0)$ as a function of $n/n_0$. The oscillatory tendencies in the figures can be attributed to the occupation of the Landau levels by the charged particles. The oscillations becomes more prominent near the higher end of the density range since the magnetic field increases with the density. The electrons, being the lightest particles, show more prominent oscillations in their particle fraction ratio from an earlier density. In case of muons, the early oscillations in their profile can be explained by their lower population density, and thus lower Fermi momentum, leading to smaller number of maximum Landau levels (Eqn. \eqref{eqn. 15}), which makes Landau quantization more prominent. For the magnetic field profile with $\mu=2{\times}10^{31} Am^2$, near the surface and outer core where density is less than $2n_0$, the field strength is of the order of $10^{16}$ G. At this field strength the protons are not affected substantially and electrons are little affected by the presence of magnetic field, as can be seen from Fig. \ref{fig.5}. Hence the particle fraction and threshold of $\Delta^-$ are also least affected. The proton population is affected very little and electrons populate less number of Landau levels when the field strength reaches magnitude of the order $10^{17}$ G near density $\sim 5n_0$. Then electron fraction is increased leading to later appearance of $K^-$ and $\delta Y_i<1$ for $K^-$ in the subsequent densities. Consequently, the threshold densities of  $\Delta^+,~\Xi^0$ and $\Delta^{++}$ changes and their populations are affected due to interplay of baryon number conservation and charge neutrality condition, as seen from Fig. \ref{fig.5}. The threshold densities of various particles are given in Table \ref{table 3}. For the magnetic field profile with $\mu=1.5{\times}10^{32} Am^2$, the pattern is similar but there are some differences to be noted as evident from Fig. \ref{fig.6}. The oscillations for $\delta Y_i$ of protons ($p$) and electrons ($e^-$) at the initial densities ($< 1n_0$) are noticeable. This is due to the low population of $e^-$ and $p$, and the high magnitude of magnetic field, of the order $10^{17}$ G, for this profile near the surface and outer core. Here, we can also see that the $e^-$ and $p$ populations oscillate in unison. This is because they are the only charged particles at this density range and, thus, to maintain charge neutrality they must increase or decrease similarly. %The oscillations for $e^-$ and $p$ dies down around density $\sim 1n_0$. The oscillations in $\delta Y_i$ for $e^-$ increases again at higher densities and becomes quite large for densities $>3n_0$ where the field strength becomes close to $10^{18}$ G. 
Similar to the case for $\mu=2{\times}10^{31} Am^2$, the electron population increases substantially when the field strength reaches around $10^{18}$ G near density $3n_0$. %in Fig. \ref{fig.5}, 
This results in the later appearance of $K^-$ and $\delta Y_{K^-} < 1$. The amplitudes of the oscillations in $\delta Y_{e^-}$ being larger compared to the case of field profile with $\mu=2{\times}10^{31} Am^2$, the threshold density for appearance of $K^-$ is pushed towards an even higher density in this case  and $\delta Y_{K^-}$ is also smaller in this case. The oscillations of the charged particles are, in general, significantly larger in Fig. \ref{fig.6} than in Fig. \ref{fig.5}, as expected. Similar to the case for field profile with $\mu= 2{\times}10^{31} Am^2$, the threshold densities and the particle populations for $\Delta^+$, $\Xi^0$ and $\Delta^{++}$ are also altered in the case for field profile with $\mu=1.5{\times}10^{32} Am^2$ but the change is greater for the latter case, as can seen from Table \ref{table 3}. We also notice that $\delta Y_{\Xi^0}$ for DD-ME2 parametrization does not appear in  Fig. \ref{fig.6} as it is very low and outside the range of the plot. This is due to it's appearance being delayed because of the presence of strong magnetic field. 

In Fig. \ref{fig.7}, we illustrate the relationship of the Dirac effective mass for nucleons as a function of of $n/n_0$. We observe that deviative features start appearing in the figures with the onset of $\Delta^-$ around density $\sim 1.6n_0$.
\textcolor{black}{This oscillating behaviour is associated with the Landau quantization, which in turn will affect the matter properties viz. specific heat, mean-free path of baryons, thermal conductivity to name a few.}

%\textcolor{red}{\textbf{Add about fig.-\ref{fig.10}...}}
\textcolor{black}{In Fig. \ref{fig.10}, we illustrate the matter pressure density as a function of electron chemical potential ($\mu_e$) and neutron chemical potential ($\mu_n$). We observe that $\mu_e$ increases initially and then starts to decrease from a point which corresponds to the appearance of $\Delta^-$ particles. This is attributed to the replacement of electrons by $\Delta^-$ in maintaining charge neutrality. We also note that the slope of the plot softens slightly after the appearance of $K^-$ condensates. This happens because (anti)kaon condensates being $s$-wave Bose condensates do not contribute to the matter pressure, after they replace baryons in the matter composition as favoured from energy argument point of view.}

\begin{figure*}
    \centering
\includegraphics[width=16.5cm,keepaspectratio ]{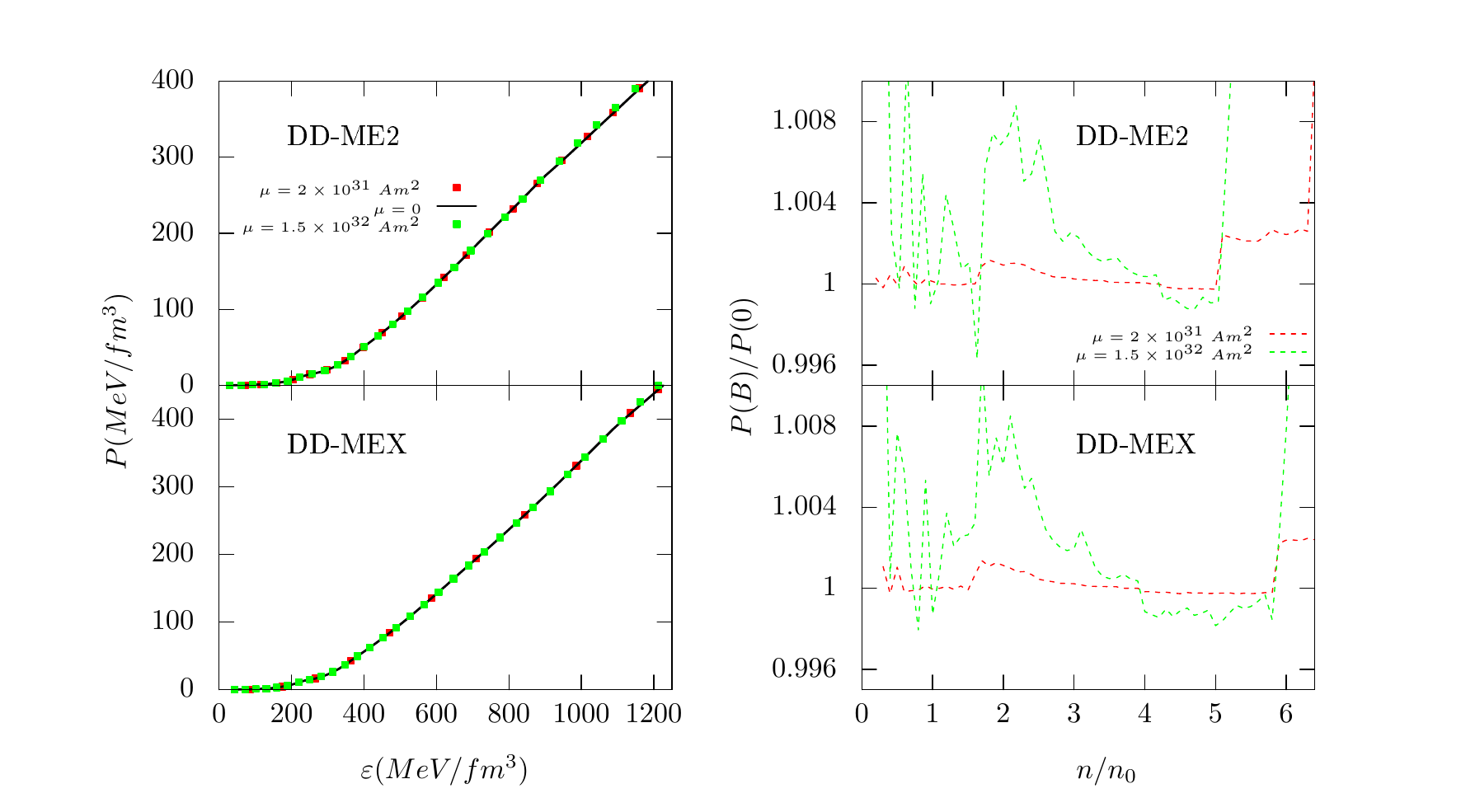}
    \caption{Left panel: EoS for NS matter in the presence of magnetic fields with $\mu=1.5{\times}10^{32}Am^2$ and $\mu=2{\times}10^{31}Am^2$, and also for NS matter in the absence of magnetic field ($\mu=0$). Right panel: Ratio of pressure density in the presence of the same two magnetic fields, $P(B)$, to pressure density in the absence of magnetic field, $P(0)$, as a function of the total number density normalized to $n_0$ ($n/n_0$). The plots are given for DD-ME2 (upper plots) and DD-MEX (lower plots) parametrizations. The matter composition considered is $N\bar{K}Y\Delta$.}
    \label{fig.8}
\end{figure*}

\textcolor{black}{Due to the strong magnetic fields}, in the higher density regime the matter EOS stiffens. Although it is not evident from the left panel of Fig. \ref{fig.8} which illustrates the EOS for $N\bar{K} Y\Delta$ composition for the two parametrizations( DD-ME2 and DD-MEX) and for the two magnetic field profiles (with $\mu=2{\times}10^{31} Am^2$ and $\mu=1.5{\times}10^{32} Am^2$). However, the feature is evident from the right panel of Fig. \ref{fig.8} which shows the ratio of pressure density in presence of magnetic field $P(B)$, to pressure density without magnetic field, $P(0)$, as a function of total number density fraction, $n/n_0$. 
\textcolor{black}{The right panel manifests the minute effects due to Landau quantization in dense matter.}
The oscillations in this figure, as in the case of Fig. \ref{fig.5} and Fig. \ref{fig.6}, can be attributed to the occupation of the Landau levels by the charged particles at high value of magnetic field.

\begin{figure*}
    \centering
\includegraphics[width=16.5cm,keepaspectratio ]{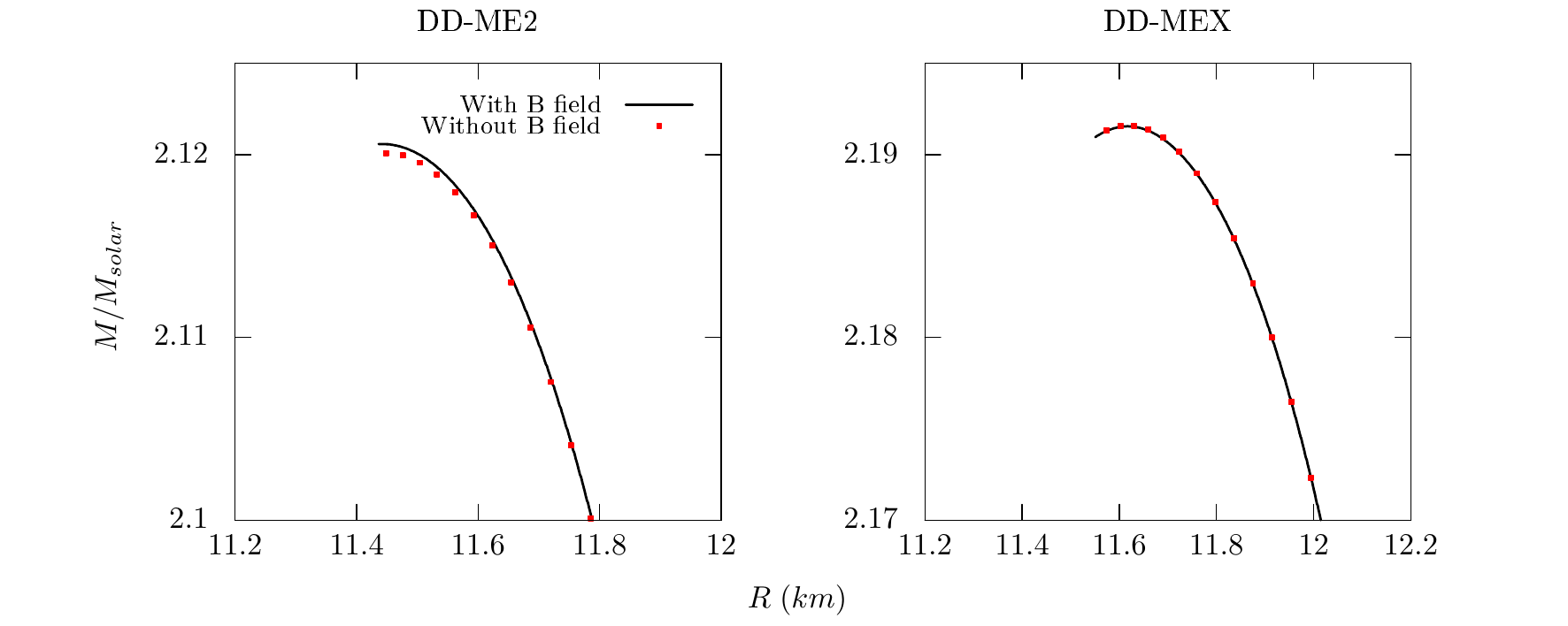}
    \caption{Comparison of the closeup of the mass-radius relations in the region of the maximum mass. The plots are given for NS matter with magnetic field ($\mu=2{\times}10^{31}Am^2$) and without magnetic field, and for the two parametizations: DD-ME2 (left panel) and DD-MEX (right panel). The matter composition considered is $N\bar{K}Y\Delta$. }
    \label{fig.9}
\end{figure*}

Now, so far inferred maximum limit of surface magnetic field strength from the magnetar observations is of the order of $\sim~10^{16}$ G. So we consider the field profile with $\mu=2{\times}10^{31}Am^2$ which gives the surface field strength $\sim~10^{16}$ G. With this profile the maximum field strength within the star remains below the order of $10^{17}$ G. Hence, for this field profile, the solution of TOV equations for the star structure can be taken as a good approximation.
We show the effect of the magnetic field on the maximum mass of NS with $N \bar{K} Y \Delta$ composition in Fig. \ref{fig.9}. We observe a small increase ($\sim 0.05\%$) in maximum mass which is visible for the case of DD-ME2 parametrization. This is the consequence of the stiffening of matter in presence of magnetic field due to late appearance of $K^-$.%and we observe no significant for the case of DD-MEX. However, in Fig. \ref{fig.9}, we only provide the M-R relation for NS matter in the presence of magnetic field profile with $\mu=2{\times}10^{31}Am^2$ since for stronger magnetic field profiles the maximum field strength exceeds $10^{17}$ G and we can no longer use Eqn. \ref{eqn.33}. }

\section{Summary and Conclusion}\label{sec:summary}
Recent observations of massive NSs \cite{_zel_2010, 2010Natur.467.1081D, 2020NatAs...4...72C, 2013Sci...340..448A, 2022ApJ...934L..17R} suggest the existence of matter \textcolor{black}{to be} at densities above $2n_0$ inside the core of the massive stars. At such high densities it is quite possible for exotic degrees of freedom of matter to appear. In this scenario, the possibility of appearance of strange and non-strange heavier baryons  \cite{1960SvA.....4..187A, LI2018234,Li_2019,PhysRevD.89.043014}, kaons \cite{PhysRevC.90.015801, PhysRevC.60.025803,PhysRevC.63.035802,PRAKASH19971,PhysRevC.53.1416,PhysRevC.64.055805,Malik2021NewEO,PhysRevD.102.123007, PhysRevC.77.045804} \textcolor{black}{and strange quark matter (SQM)} \cite{1999JPhG...25..195W, 1986A&A...160..121H, 2004JPhG...30S.471T} inside the core of the NSs, and its consequences on the NS observables, are of great interest and are discussed \textcolor{black}{in the above cited literature and many others.}
The effect of strong magnetic field on highly dense matter with kaons and in absence of heavier baryons has been discussed previously \cite{PhysRevC.77.045804,2022IJMPE..3150050K}. In our present work, we discussed the effect of presence of strong magnetic field on highly dense NS matter with all possible \textcolor{black}{baryonic exotic degrees of freedom, viz. hyperons and $\Delta$-resonances, and (anti)kaon condensates,} in view of the existence of magnetars with high surface magnetic field.  The inferred surface field strength from the magnetar observations is in the range $10^{13-16}$ G. The field strength inside the NSs can not be inferred from any observation as of yet but can be estimated theoretically from the solution of combined Einstein-Maxwell field equations. Hence, we considered a model magnetic field profile, as a function of baryon chemical potential inside the NS, which is poloidal in nature and satisfies the Einstein-Maxwell field equations. With this profile, the field strength gradually increases towards the centre of the star and at the centre the strength is $\sim 1$ order higher compared to that at the surface.
\textcolor{black}{Furthermore, the (anti)kaon condensate appears at high density compared to the threshhold density of deconfinemnet to SQM. Hence, if we consider a hybrid star (HS) configuration, the occurence of (anti)kaon condensate is very unlikely inside the core of a HS. Further investigation in this particular aspect has not been explored and is beyond scope of this work.}

For our discussion, we considered the model of matter within DD-RMF model with two parametrizations, DD-ME2 and DD-MEX, which are compatible with the astrophysical observations for NSs, with matter composed of $N \bar{K} Y \Delta$. The variation of magnetic field with the matter density is more or less the same with these two parametrizations for the considered field profile. The presence of magnetic field pushes the threshold for appearance of $K^-$ to a higher density. Consequently, at higher density regime, the matter stiffens compared to the case without magnetic field and this effect is more in the case of DD-ME2 parametrization than the DD-MEX parametrization. This leads to increase in the maximum attainable mass, compared to the case without magnetic field, which is also more prominent for the DD-ME2 parametrization. 

\begin{acknowledgments} 
The authors thank the anonymous referee for constructive comments which enhanced the 	quality of the manuscript. M.S. acknowledges the funding support from Science and Engineering Research Board, Department of Science and Technology, Government of India through Project No. CRG/2022/000069. V.B.T. acknowledges the funding support from a grant of the Ministry of Research, Innovation and Digitization through Project No. P4-ID-PCE-2020-0293. 
\end{acknowledgments}

%\newpage
\bibliography{references}
\end{document}